\documentclass[sigconf]{acmart}
\usepackage{algorithm}
\usepackage{algorithmic}
\usepackage{subcaption}
\usepackage{soul}
\usepackage{tabu}
\usepackage{booktabs}

\AtBeginDocument{%
	\providecommand\BibTeX{{%
			\normalfont B\kern-0.5em{\scshape i\kern-0.25em b}\kern-0.8em\TeX}}}

\acmConference[]{}{}{}
\acmBooktitle{}
\acmPrice{}
\acmISBN{}

\setcopyright{none}
\renewcommand\footnotetextcopyrightpermission[1]{}

\begin{document}
	
\title{\textbf{iDROP:} Robust Localization for \textbf{I}ndoor Navigation of \textbf{Dr}ones with \textbf{O}ptimized Beacon \textbf{P}lacement}

\author{Alireza Famili, Angelos Stavrou, Haining Wang, Jung-Min (Jerry) Park} 
\affiliation{ 
	\institution{Department of Electrical and Computer Engineering}
	\institution{Virginia Tech}
}
\email{{afamili,angelos,hnw,jungmin}@vt.edu}

\renewcommand{\shortauthors}{Famili and Stavrou, et al.}

	\begin{abstract}
		Drones in many applications need the ability to fly fully or partially autonomously to accomplish their mission. To allow these fully/partially autonomous flights, first, the drone needs to be able to locate itself constantly. Then the navigation command signal would be generated and passed on to the controller unit of the drone. In this paper, we propose a localization scheme for drones called iDROP (Robust Localization for \textbf{I}ndoor Navigation of \textbf{Dr}ones with \textbf{O}ptimized Beacon \textbf{P}lacement) that is specifically devised for GPS-denied environments (e.g., indoor spaces). Instead of GPS signals, iDROP relies on speaker-generated ultrasonic acoustic signals to enable a drone to estimate its location. In general, localization error is due to two factors: the ranging error and the error induced by relative geometry between the transmitters and the receiver. iDROP mitigates these two types of errors and provides a high-precision three-dimensional localization scheme for drones. iDROP employs a waveform that is robust against multi-path fading. Moreover, by placing beacons in optimal locations, it reduces the localization error induced by the relative geometry between the transmitters and the receiver.  
	\end{abstract}
	

	\maketitle
	
	\section{Introduction} \label{Introduction}
	\noindent Over the past few decades, the global drone industry has expanded exponentially and the number of applications in which drones play a significant part, both in indoor and outdoor environments, has subsequently increased. There is a wide range of indoor drone applications nowadays, ranging from recreational use to life-saving matters. Examples include reconnaissance inside nuclear power plants, helping firefighters to locate people inside burning buildings, security surveillance inside large warehouses, etc.     
	
	Drones must have fully or partially autonomous flying capability in most of the above applications to perform their task successfully. To allow these autonomous flights, first, the drone needs to localize itself constantly. Then the navigation command signal would be generated and passed on to the controller unit of the drone according to the current position and ultimate destination. In outdoor environments, drones can easily use GPS signals for self-localization; however, such an approach is not feasible in indoor spaces or GPS-denied areas.
	
	In the absence of the GPS, vision-based methods are widely used for localization and navigation of drones (e.g., \cite{Novel_Visual_Odometry}). However, the accuracy of current vision-based approaches is usually limited due to the drone's vibration during flight. In addition, the accuracy can degrade further in vision-impaired environments (e.g., low light environments). Moreover, vision-based methods are expensive both in terms of hardware price and computational complexity.  
	
	In addition to vision-based approaches, ranging-based methods are commonly deployed for indoor localization. In this category, the localization is based on the received signal information. RF-based localization (e.g.,\cite{Freq_Hopping_WiFi}) and localization based on acoustic signals (e.g., \cite{ROLATIN}) are the examples for this group. Besides the characteristic of the received signal, the arrangement of the beacons plays an important role in the localization accuracy in this group. 
	
	In this paper, we propose \emph{iDROP (Robust Localization for \textbf{I}ndoor Navigation of \textbf{Dr}ones with \textbf{O}ptimized Beacon \textbf{P}lacement)}, a three-dimensional localization scheme for drones in GPS-denied environments. iDROP uses ultrasonic acoustic-based signals for localization. We claim that acoustic signals have some advantages over the localization schemes based on RF signals. Compared to RF signals, the significantly slower propagation speed of the acoustic signals allows for higher accuracy of localization. In addition, RF signals can penetrate through walls and ceilings, further degrading the accuracy of the localization. iDROP uses high-frequency acoustic signals, known as \emph{ultrasounds}, to prevent any interference with human-generated or drone's propeller noise. Moreover, iDROP develops an optimization framework to find the optimal placement for the ultrasound transmitter beacons.
	Following is a summary of our contributions.
	
	$\bullet$ We propose iDROP, a three-dimensional localization scheme for drones in GPS-denied environments which is robust against noise and multi-path fading and provides location estimation with high accuracy.
	
	$\bullet$ iDROP uses the hybrid \textit{Frequency Hopping Code Division Multiple Access (FH-CDMA)} as the communication scheme to maximize robustness to noise and multi-path fading and to facilitate signal separation at the receiver.
	
	$\bullet$ iDROP develops an optimization framework to reduce the height estimation error due to the relative geometry between transmitters and receivers.
	
	$\bullet$ By leveraging the code division multiple access techniques, iDROP reduces the communication link used for navigation commands by placing the receiver on-board and transmitter beacons in the room.
	
	$\bullet$ Our simulation and experimental results indicate that iDROP's localization error is less than $1.5$ centimeters in three-dimensional space.   
	
	The rest of this paper is organized as follows. In the next section, we review some of the related works in this area. Then, in section \ref{Robust FH-CDMA Localization} which is the first core of our paper, we fully explain the first stage of our localization scheme for drones in no-GPS environments. In section \ref{Preliminary}, we describe the preliminary simulation setup and bring the results of localization with a random beacon placement similar to what they have in \cite{ROLATIN}. We evaluate the performance in this section and show that improving the localization accuracy by lessening the ranging error is not sufficient for a high-accuracy localization scheme. Next, in section \ref{Enhancing the Accuracy of Localization}, which is the second core of our paper, we thoroughly explain the methods we used to bring further enhancement to our localization scheme. Then, in section \ref{Experimental Tests and Results}, we describe our experimental testbed followed by the results and evaluation of our proposed scheme. Finally, we conclude our work in section \ref{Conclusion}.
	
	\section{Related Work} \label{Related Work}
	Our work is related to the following research studies: (i) indoor localization, (ii) beacon placement optimization, and (iii) autonomous navigation for drones in the absence (or lack) of GPS signals.
	
	Localizing a target in indoor environments without the GPS signal has been a topic of interest. Ranging-based methods are of the most well-known approaches for indoor localization. In this category, RF, acoustic, or ultrasound signals are deployed to find the distance between the beacons and the target. Then, by using the distance between the target and several beacons, the target's position is estimated by leveraging some techniques such as lateration or angulation \cite{ROLATIN, MobiSys_Follow_Me_Drone, ToneTrack, Robust_Broadband}. 
	
	Beacon placement optimization is a well-known topic to optimize the location of beacons for indoor localization purposes \cite{CMU, Efficient_Beacon_Placement, CMU_paper, Novel_Beacon_Placement, BLE_Localization_Precision_Limits} or wireless network localization \cite{Relative_location_estimation, Network_Navigation_with_Scheduling}. There are two major categories here, first optimizing the number of beacons and their location to have full coverage for the entire indoor venue and second optimizing the placement of beacons to minimize the localization error, which is due to the relative geometry between the target and beacons. For the first issue, the type of sensors plays an important role, because they have different coverage, e.g., if the system is based on low power Bluetooth sensors, the transmission would be omnidirectional and coverage is restricted just by the distance and obstacles; however, if a system uses ultrasound-based sensors, then the beam angle of the sensors also puts an additional restriction on finding the number of beacons and their placement. After the required number of beacons is fixed, the second optimization platform needs to be deployed in order to find the placement for beacons to minimize the localization error due to the relative geometry between the receiver and beacons. 
	
	For autonomous navigation of drones in the absence (or lack) of GPS signals, some well-known techniques tackle the problem. Vision-based models using different visual techniques such as visual odometry (VO), simultaneous localization and mapping (SLAM), and optical flow \cite{low_cost_solution, 6_Dimensional, Survey_UAV_navigation_GPS_denied}. The major drawback of using merely visual techniques is poor image quality due to the obstacles in a drone's flying environment, which degrade the accuracy. There are a few research papers where they tried to resolve this issue by leveraging some auxiliary techniques, e.g., some used deep neural networks in combination with visual techniques (e.g., \cite{Neural_Network}) or some gained benefits from the use of LiDAR (e.g., \cite{LiDAR}) for autonomous flying.
	
	\section{Robust Localization with Hybrid FH-CDMA Ultrasound Signals} \label{Robust FH-CDMA Localization}
	iDROP is a novel and highly accurate three-dimensional localization scheme for drones in indoor environments which deploys two steps to reduce both sources of localization error, known as the ranging error and the error due to the relative geometry between the receiver and transmitter beacons.
	
	This section thoroughly investigates how iDROP reduces the ranging error and increases the localization accuracy by making the system robust against noise and the indoor multi-path fading effect. Then, in section \ref{Enhancing the Accuracy of Localization}, we will elaborate how it minimizes the error due to the relative geometry and increases the overall accuracy of localization by optimizing the placement of beacons in the room.
	
	\subsection{Measurement Method and Technique}
	As we have discussed earlier in Sec.~\ref{Introduction}, the ranging-based localization with ultrasonic signals has some advantages over the other techniques. Hence, iDROP uses ultrasound acoustic signals for distance estimation. Well-known measurement methods are the angle of arrival (AOA), time of arrival (TOA), time difference of arrival (TDOA), and received signal strength (RSS). Techniques for location estimation are angulation, lateration, and fingerprinting. AOA methods incur high expenses in terms of both the hardware cost and the processing power because they require special antenna arrays and complicated calculations. RSS and fingerprinting are too prone to changes in real-time and are therefore neither reliable nor highly accurate. All said, iDROP uses the trilateration technique and the TOA of the received ultrasound signals for localization. Multi-path fading is one of the main challenges of relying on the TOA of the received signal. The presence of the copied version of the original signal in the receiver makes detecting the arrival time of the original signal hard or even impossible. iDROP overcomes the multi-path fading effect by proposing a hybrid FH-CDMA communication scheme for its signal transmission.
	
	\subsection{Implementation Challenges}
	In terms of placement of the transmitter(s) and receiver(s) for a ranging-based localization, there are two general scenarios, either have the receiver(s) on-board the drone and keep the transmitter(s) in the room or vice versa. The localization calculation task takes place on the receiver side of the system; therefore, not having them on-board the drone requires another communication link for sending the final location estimation to the drone. This unnecessary communication link increases the cost, slows down the whole process, and may incur additional errors, which degrades the accuracy. 
	
	Therefore, it is better to have the receiver(s) mounting on-board the drone and the transmitter(s) in the room. In this case, having one transmitter in the room and multiple receivers on-board the drone \cite{Ultrasonic_Quadrotor_2019} causes several problems. It adds extra weight to the drone, increases the power consumption, and most importantly, there is not enough space between the receivers which incurs error due to the relative geometry between the transmitter and receivers and significantly degrades the accuracy of the localization. 
	
	To overcome these challenges, iDROP mounts one receiver on-board the drone and keeps all the transmitters spatially distanced from each other in the room. However, this method raises a new challenge, the need for signal separation in the receiver. The receiver requires the capability of separately detecting the TOA of each signal transmitted from a different transmitter. To rectify this matter, iDROP deploys a code division technique. It assigns a code to the transmitted signals of each of the ultrasound transmitters in the room, i.e., transmitted signals at each transmitter are encoded using a code that is orthogonal to all other transmitters' code. Having four transmitters, iDROP generates a different orthogonal code for each transmitter using a Walsh-Hadamard matrix of size four. Data bits of each transmitter would be multiplied with one of the rows of this matrix. At the receiver side, received signals will be multiplied with all the four codes and signals from each transmitter get detected.
	
	\begin{figure}
		\centering
		\includegraphics[height=1.2in,width=3.3in,trim={0.5cm 14cm 18cm 0},clip]{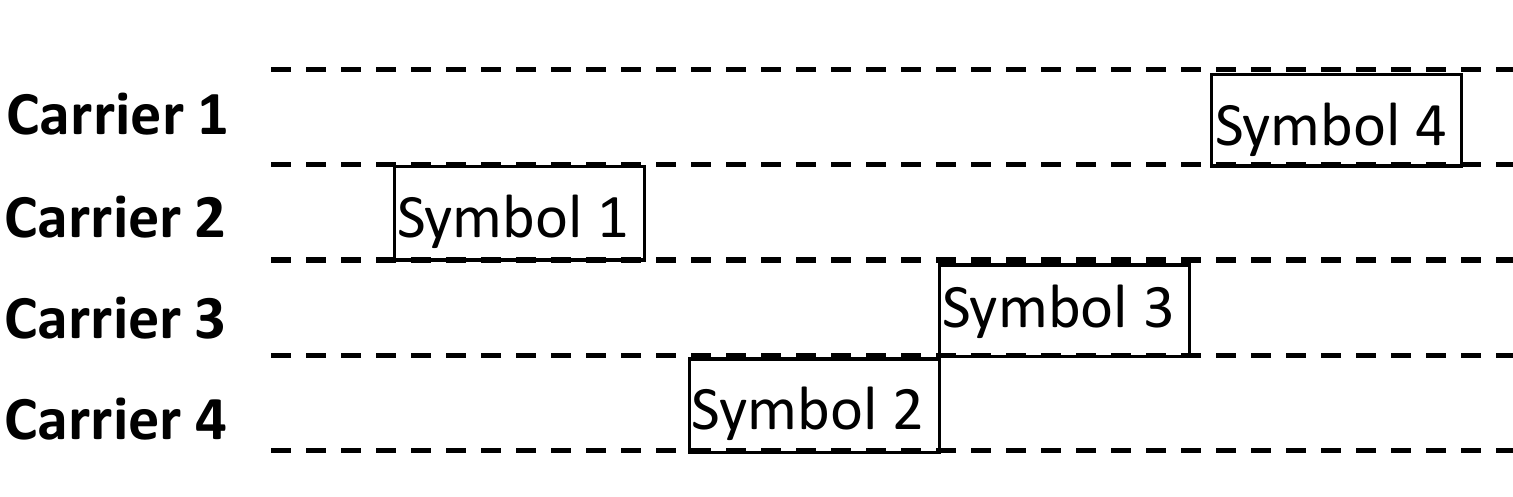} 
		\caption{This figure illustrates the hybrid FH-CDMA transmission of encoded symbols over different FH channels.}
		\label{FH-CDMA}
	\end{figure}
	
	\subsection{Hybrid FH-CDMA}
	
	iDROP deploys a hybrid FH-CDMA technique which, to the best of our knowledge, is the first time this technique has been applied for a localization purpose and it is the most desirable scheme to address both problems of multi-path and signal separation. The hybrid FH-CDMA is a communication scheme that combines two well-known techniques, the Frequency Hopping (FH) and the Code Division Multiple Access (CDMA). iDROP uses this method to rectify the challenge of signal separation in the receiver with the multiple access capability and at the same time, brings robustness against noise and the indoor multi-path fading using the frequency hopping technology.
	
	In our system, as long as we make sure that the hops happened fast enough that before the first multi-path reflection arrived at the receiver, we already hopped to another carrier frequency, then we can promise a transmission that is robust against the multi-path fading. We picked the hopping rate equal to the symbol rate, which is fast enough to avoid multi-path according to the room channel characteristic.
	
	First, at each ultrasound transmitter beacon in the room, the code assigned to that transmitter would be assigned to each data bit of that transmitter. Therefore the symbols are no longer just a bit; they are coded bits that include four bits, i.e., data-symbols of different beacons are spread with their assigned code and generated the coded symbols. Then, since the hop rate equals the symbol rate, each coded symbol is transmitted in different FH channels. Here, the coded symbols enable signal separation of different transmitters in the receiver side and the different frequency channels are for managing the multi-path fading effect of the room.
	
	Similar to \cite{Robust_Broadband}, in our scheme, the transmitting signal of the $i$-th transmitter is modulated using Binary Phase Shift Keying (BPSK) modulation, then encoded with its dedicated code, and then the coded symbols are spread using a sinusoidal signal with a variable frequency depending on the pseudo-random code which is known both in transmitter and receiver side:
	\begin{eqnarray}\label{Tx_FHSSS}
	s^{(i)}(t) = d^{(i)}\cdot c^{(i)} \cdot pT_B(t)\cdot \sin(2\pi f_mt+\phi),
	\end{eqnarray}
	where $T_B$ is the data symbol duration, $d^{(i)}\cdot c^{(i)}$ is the transmitted symbol of the $i$-th ultrasonic transmitter in the room where $d^{(i)}$ is the data bit and $c^{(i)}$ is the dedicated code to that transmitter, the rectangular pulse $pT_B$ is equal to $1$ for $0 \leq t < T_B$ and zero otherwise, and $f_m$ is the set of frequencies over which the signal hops. Then the received signal is in the form of:
	\begin{eqnarray}\label{Rx_FHSS}
	& r = \sum_{i=1}^{4} s^{(i)}(t-\tau_i) + \mathcal{M} + \mathcal{N},\nonumber
	\end{eqnarray}
	where $\tau_i$ is the propagation delay from the $i$-th transmitter to the receiver on-board drone that we are using for calculating the distance, $\mathcal{N}$ is the overall Gaussian noise, and $\mathcal{M}$ is the summation of all the multi-path fading effects:
	\begin{eqnarray}\label{Multi-path}
	\mathcal{M} =\sum_{i=1}^{4} \sum_{j=1}^{N} \alpha_{ij}\cdot s^{(i)}(t-\tau_{ij}),
	\end{eqnarray}
	where $\alpha_{ij}$ is the attenuation of path $j$ for the $i$-th transmitter, $\tau_{ij}$ is the time delay of the path $j$ for the $i$-th transmitter, and $N$ is the number of multi-path signals. Multi-path is a big issue for indoor environments and we are using the frequency hopping technique to help overcome this effect. As long as we make sure that at each transmitter, the speed of hopping is faster than the time delay of all the multi-path signals corresponding to that transmitter ($\tau_{ij}$), then before the arrival of any of the reflected signals, we already have changed the frequency and different paths will not interfere with the original signal.
	
	By ensuring that multi-path effects are eliminated using different FH-channels, the received signal would be only the delayed time of the transmitted signal plus noise:
	\begin{eqnarray}
	r  = \sum_{i=1}^{4} s^{(i)}(t-\tau_i) + \mathcal{N}.
	\end{eqnarray}
	By multiplying the received signal in each code related to each transmitter, the received signal from the $i$-th transmitter in the receiver would be in the form of:
	\begin{eqnarray}\label{Rx_FHSS_without_multipath}
	r^{(i)}  = d^{(i)}\cdot pT_B(t-\tau_i)\cdot \sin(2\pi f_m(t-\tau_i)+\phi) + \mathcal{N}.
	\end{eqnarray}
	Therefore, by implementing a cross-correlation between the received signal and the known transmitted signal (the one without the time delay) and detecting the sample at which the peak occurs, the distance is calculated as the following:
	\begin{eqnarray}\label{Distance_Calculation_2}
	d = \frac{n_{samples}}{f_s}\cdot c_{sound},
	\end{eqnarray}
	where $n_{samples}$ is the sample number of the maximum peak, $f_s$ is the sampling frequency, and $c_{sound}$ is the speed of sound. 
	
	\subsection{Three-dimensional Localization}
	After having successfully measured the distance between an ultrasonic transmitter and the receiver, the next step is the three-dimensional localization of the receiver. To localize an object in two dimensions using trilateration, at least distances between the object and three sources are needed. Similarly, in three-dimensional localization, it is required to have the distance between the object and at least four sources to localize the object uniquely. Let's denote the distance between the receiver and the $i$-th transmitter as $d_i$. Also, the position of the receiver is $[x \ y \ z]^T$ (which in fact is the position of the drone) and the position of the $i$-th transmitter denotes as $[x_i \ y_i \ z_i]^T$. Then using trilateration rules, we have:
	\begin{eqnarray} \nonumber
	&(x_1-x)^2+(y_1-y)^2+(z_1-z)^2 = d_1^2\\ \nonumber
	&(x_2-x)^2+(y_2-y)^2+(z_2-z)^2 = d_2^2\\ \nonumber
	&\vdots\\
	&(x_n-x)^2+(y_n-y)^2+(z_n-z)^2 = d_n^2 
	\end{eqnarray}
	We can then simplify these quadratic equations and write them down in the form of $\textbf{A}\textbf{x} = \textbf{b}$ where $\textbf{A}$ and $\textbf{b}$ are equal to:
	\begin{small}
		\begin{eqnarray}
		\textbf{A} & = & \begin{bmatrix}
		2(x_n-x_1) & 2(y_n-y_1) & 2(z_n-z_1) \\
		2(x_n-x_2) & 2(y_n-y_2) & 2(z_n-z_2)\\
		\vdots     & \vdots     & \vdots\\
		2(x_n-x_{n-1}) & 2(y_n-y_{n-1}) & 2(z_n-z_{n-1})\\
		\end{bmatrix},\nonumber \\
		\textbf{b} & = & \begin{bmatrix}
		d_1^2 - d_n^2 - x_1^2 -y_1^2 -z_1^2 + x_n^2 + y_n^2 + z_n^2 \\
		d_2^2 - d_n^2 - x_2^2 -y_2^2 -z_2^2 + x_n^2 + y_n^2 + z_2^2 \\
		\vdots \\
		d_{n-1}^2 - d_n^2 - x_{n-1}^2 -y_{n-1}^2 -z_{n-1}^2 + x_n^2 + y_n^2 + z_n^2
		\end{bmatrix}.\nonumber
		\end{eqnarray}
	\end{small}
	The vector $\textbf{x} = [x \ y \ z]^T$ which includes the coordinate of the target drone would be: $\textbf{x} = (\textbf{A}^T\textbf{A})^{-1}\textbf{A}^T\textbf{b}$. 
	
	\subsection{Summary}
	In this section, we have elaborated how the FH-CDMA localization stage of the iDROP solves the challenge of signal separation in the receiver side in addition to providing robustness against the noise and the multi-path fading effect of an indoor environment. By encoding the transmitting signals of each ultrasound transmitter with a code generated using a Walsh-Hadamard matrix of size four, iDROP guarantees the capability of signal separation at the receiver side. Then, for robust transmission against noise and multi-path, iDROP uses different frequency hops to transmit encoded symbols of each transmitter. Therefore, every data symbol is spread with a complete orthogonal code while successive symbols are transmitted in different frequency hopped channels. Fig.~\ref{FH-CDMA} illustrates this better.
	
	\section{Preliminary Simulation Results} \label{Preliminary}
	This section evaluates the performance of the proposed FH-CDMA localization and shows the localization error for $X$, $Y$, and $Z$ axis separately. First, in \ref{Preliminary_A}, we describe the test setup, and then, we show the simulation result in \ref{Preliminary_B}. 
	
	\subsection{Preliminary Simulation Setup} \label{Preliminary_A}
	The performance of the localization scheme proposed in section \ref{Robust FH-CDMA Localization} is assessed by implementing simulation in MATLAB. Similar to \cite{ROLATIN}, we locate the transmitters at the $(x,y,z)$ coordinates equal to $(2.5,0,1.5)$, $(5,2.5,2.5)$, $(2.5,5,2)$, and $(0,5,3)$ where all the numbers are in the meter unit. To better observe how we conduct the simulation, we break it down into three sub-systems as follows.
	\subsubsection{Transmitter}
	The transmitter sub-system, which in fact is the ultrasonic transmitters at the known positions in the room, generates the desired FH-CDMA signals. We used signals in the frequency range of $20$ KHz to $50$ KHz for two reasons. First, to avoid inciting excessive audible noise or facing interference from human-generated voice, we pick frequencies over $20$ KHz to avoid overlapping with the human voice frequency range. Any overlaps may cause interference and result in degrading the performance. On the other hand, according to the Nyquist theorem, the sampling rate needs to be at least twice the maximum frequency to avoid aliasing; hence, if the system works in the frequency range of $20$ KHz to $50$ KHz, then the sampling frequency needs to be at least $100$ KHz. To avoid the cost of processing and equipment, dealing with high frequency is not suitable; therefore, we do not transmit above $50$ KHz, which means that the sampling rate simply could be $100$ KHz or more. 
	To generate the FH-CDMA waveform successfully, iDROP uses $6$ different frequency hops with $5$ KHz bandwidth dedicated to each hop. Also, it assigns a code to each transmitter, so every data bit of each transmitter would first be multiplied with the code and then transmitted via one of the six frequency hops centered at frequencies $22.5$ KHz, $27.5$ KHz, $32.5$ KHz, $37.5$ KHz, $42.5$ KHz, and $47.5$ KHz. The codes are orthogonal to each other and made by a Walsh-Hadamard matrix of size $4$. At each hop, one data symbol which already been multiplied by its code would be transmitted, so the hop rate is equal to the data bit rate (actual data bit rate before multiplying by the code), which is fast enough to mitigate the multi-path fading effect of the indoor environment. Although a sampling rate of $100$ KHz would be enough for our simulation, we picked sampling frequency ($f_s$) equal to $340$ KHz to make sure it would be large enough to avoid aliasing and it also helps to simplify some of our calculations. Since the throughput is not essential in our case, we use BPSK modulation which does not have a good transmission rate, but is highly robust against noise.
	\subsubsection{Channel}
	The channel sub-system adds white Gaussian noise (AWGN) and simulates the multi-path fading of indoor environments. The drone’s movement is assumed to be restricted to a rectangular room with dimensions of $5$~m~$\times$~$5$~m~$\times$~$4$~m. We use a Rayleigh channel with several paths for simulating the multi-path fading effect due to the reflection of the signal from walls, floor, ceiling, and other objects and obstacles in the room. The Rayleigh channel parameters that we set for our simulations are the sample rate, maximum Doppler-shift, number of different paths, delay of each path, and average path gain. We set all these parameters with respect to a typical indoor room environment. Any possible effects of noise and fading that may negatively impact the accurate detection of the TOA of the original signal are simulated using this Rayleigh channel and the AWGN added to the signal.
	\subsubsection{Receiver}
	The receiver sub-system, which is the ultrasound receiver on-board the drone, separates the signals from different transmitters by multiplying them into the transmitters' codes and demodulating the frequency hopped signals. Then, it cross-correlates the received signal with the original version and finds the bit which makes the peak in the cross-correlation and using that, it estimates the distance from each transmitter to the drone using: 
	$d = n_{samples}\times c_{sound}/f_s$, where $n_{samples}$ is the sample number that the maximum cross-correlation occurs and $f_s$ is the sampling frequency.
	
	\subsection{Results and Motivation} \label{Preliminary_B}
	Here, we present the results of our preliminary simulations using the simulation setup we described. We assess the performance of the FH-CDMA localization by calculating the error between the actual position of the drone and our estimated position. We have used a Monte Carlo method with an adequately large number of iterations for each simulation.
	
	\begin{figure}
		\centering
		\includegraphics[width=\linewidth]{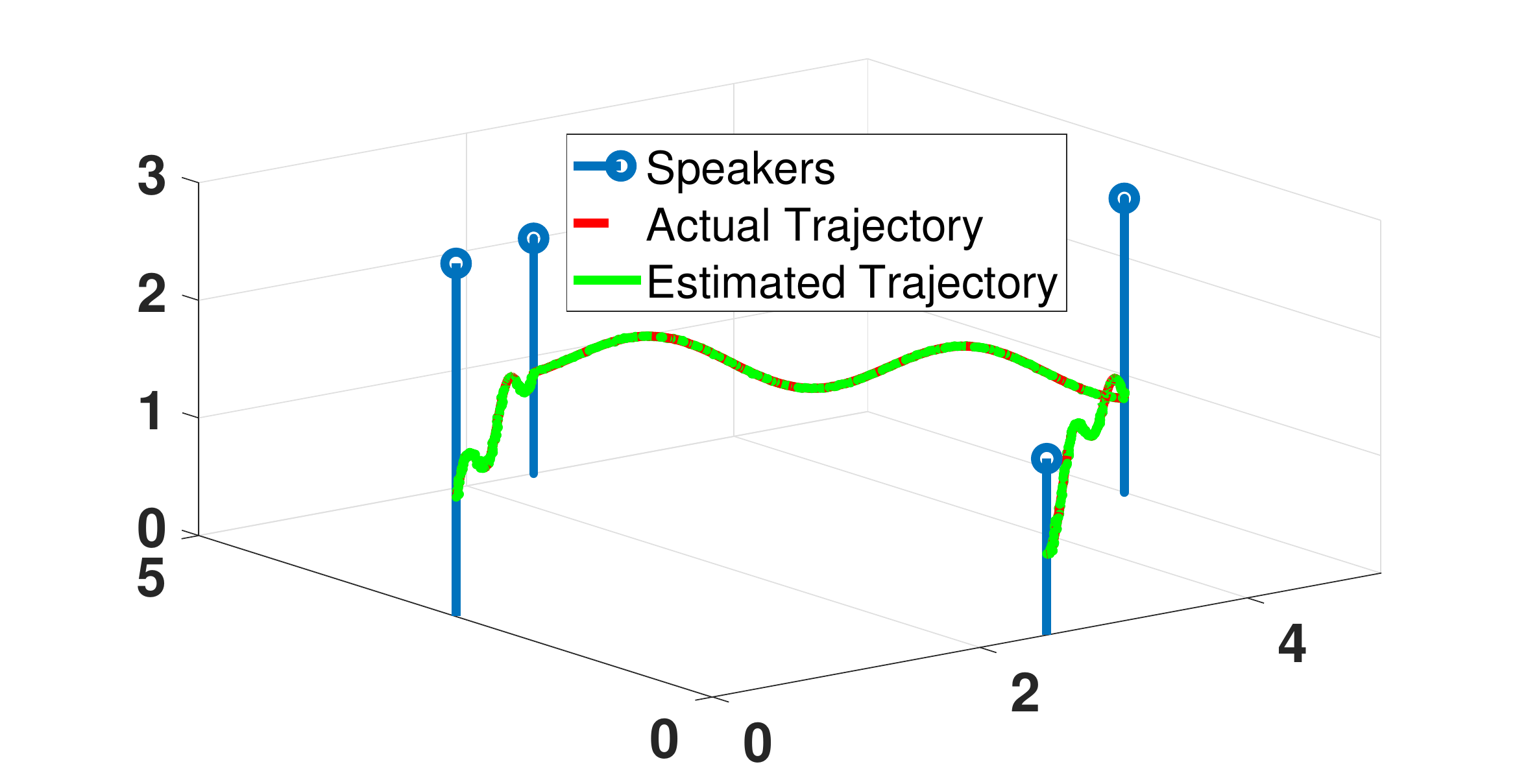} 
		\caption{Representation of the ultrasound speaker transmitter placement in the room and a comparison between the estimated trajectory of a drone and its actual trajectory. Other objects in the room are not shown in this plot to clearly show the drone's trajectory.}
		\label{trajectory}
	\end{figure}
	
	In Fig.~\ref{trajectory}, we show a drone's actual trajectory as well as the estimated trajectory using just FH-CDMA localization (the first stage of the iDROP). The locations of the ultrasound speaker transmitters are also indicated in this figure. The actual and estimated trajectories seem to overlap perfectly because the localization estimation error is significantly small relative to the room's dimensions. All the other obstacles and objects in the room; including a table, several chairs, glass windows, etc., are not shown in the figure to clearly show the drone's flight trajectory.
	
	\begin{figure}
		\centering
		\includegraphics[width=\linewidth]{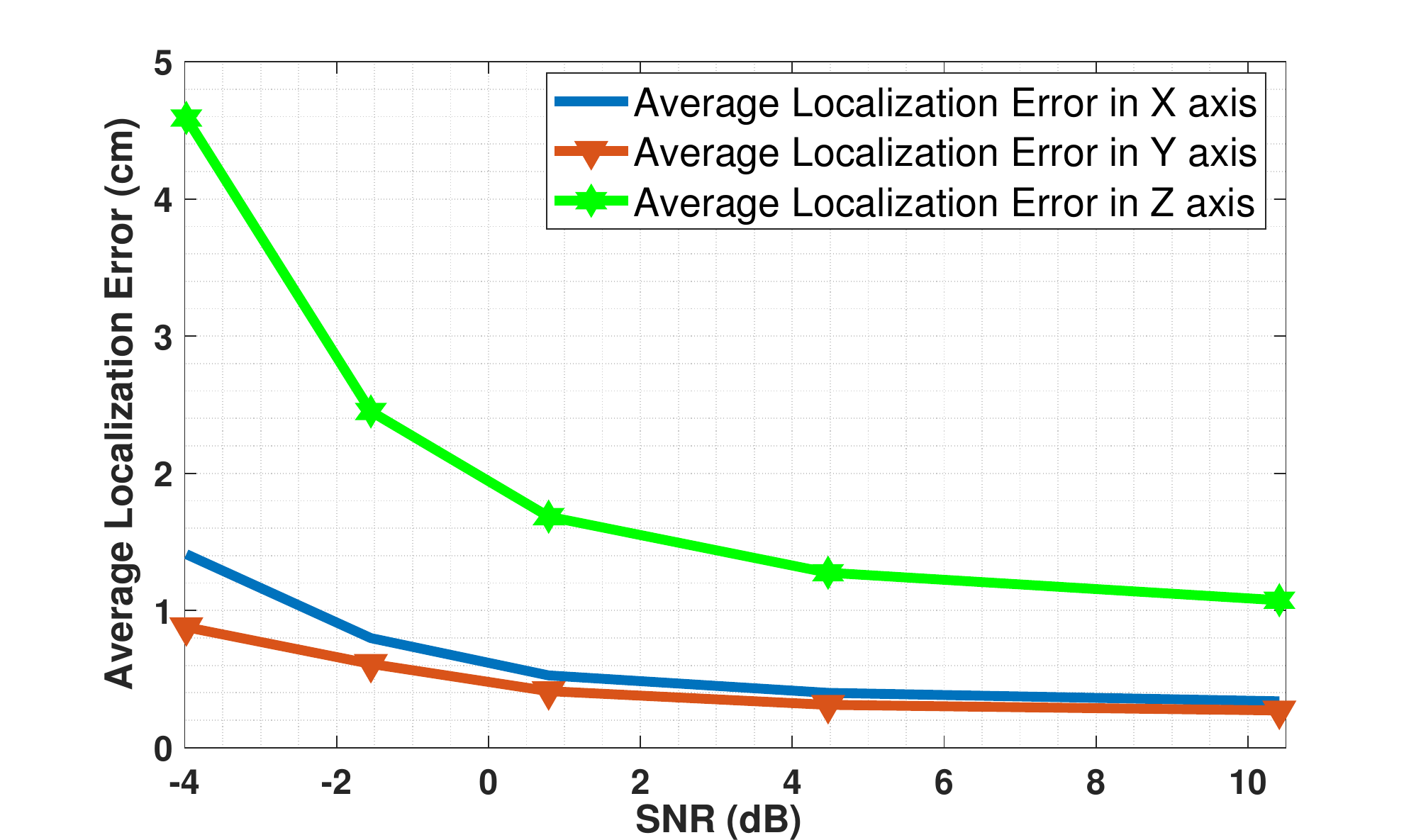} 
		\caption{Average Localization error vs. SNR (dB).}
		\label{SNR}
	\end{figure}
	
	Fig.~\ref{SNR} shows the relationship between FH-CDMA localization performance and the signal-to-noise ratio (SNR) of the signal received by the ultrasound receiver. The localization error is inversely proportional to the SNR value of the signal, as expected. In the figure, note that the $Z$-axis localization error is much greater than that of the $X$ or $Y$ axis at any given SNR.
	
	By conducting more simulations with different drone trajectories, we observed that the error of $Z$-axis localization is always drastically more than the $X-Y$ plane localization error. This is despite the fact that all the X, Y, and Z axes should have almost similar errors because they face a similar channel. This observation motivated us to further investigate the localization error factors and propose a scheme to improve the $Z$-axis localization error.
	
	\section{Enhancing the Accuracy of Localization} \label{Enhancing the Accuracy of Localization}
	As we saw in the previous section, the localization error for the $Z$-axis is much greater than that of the $X$ or $Y$ axis. This is due to the relative geometry between the transmitter beacons and the target receiver. In general, localization error for ranging-based localization is originated from two sources, first the error in estimating the distance between the target and each of the beacons, known as ranging error, and the other resulted from the relative geometry between the target and beacons.
	
	In section \ref{Robust FH-CDMA Localization}, we showed how iDROP lessens the ranging error by deploying the hybrid FH-CDMA communication scheme for distance estimation and making the scheme robust against the noise and the indoor multi-path fading effect. This section shows how iDROP copes with the error induced by the relative geometry between the drone and ultrasound transmitter beacons. To the best of our knowledge, iDROP is the first scheme that provides solutions to mitigate both the ranging and the geometry-related errors and proposes highly accurate localization for drones in indoor environments.
	
	\subsection{Dilution of Precision}
	A useful metric for measuring the localization accuracy is the Cramer-Rao Bound (CRB) which is the lower bound on the location variance that can be achieved using an unbiased location estimator \cite{CMU}. In \cite{CMU}, Rajagopal showed that for a $2$D trilateration system with an unbiased estimator, under the assumption that the range measurements are independent and have zero-mean additive Gaussian noise with constant variance $\sigma^2_r$, the CRB variance of the positional error $\sigma^2(r)$ at position $r$, as defined by $\sigma^2(r) = \sigma^2_x(r) + \sigma^2_y(r)$ is given by:
	\begin{eqnarray}
	\sigma(r) = \sigma_r \times \sqrt{\frac{N_b}{\sum_{i=1}^{N_b-1}\sum_{j=i+1}^{N_b}A_{ij}}}, \nonumber
	\end{eqnarray}
	where $N_b$ is the number of beacons, $A_{ij} = |\sin(\theta_i - \theta_j)|$, $\theta_i$ is the angle between $b_i$ and $r$, and $b_i$ is the $i$-th beacon.
	
	This shows that the localization error is a multiplication of the ranging measurement error with another term, which is the function of the number of beacons and the angle between beacons and the target object. In satellite calculations, this function is called Geometry Dilution of Precision (GDOP), therefore: $\sigma(r) = \sigma_r \times GDOP$. As CRB is directly proportional to the GDOP, we can consider GDOP as a reasonable guideline to measure the localization accuracy \cite{CMU, Relative_location_estimation, Cellular_Mobile_Estimation, Cooperative_localization}.
	
	In general, for $3$D localization of an object at $(x,y,z)$ using ultrasound beacons, we have:
	
	\begin{eqnarray}
	GDOP \cdot \sigma_r = \sqrt{Var(x)+Var(y)+Var(z)+Var(c\tau)}, \nonumber
	\end{eqnarray}
	where $c$ here is the speed of sound and $\tau$ is the receiver's clock offset. In our simulations, we assume that the transmitter and receiver use the same clock, and hence we set the timing offset to zero. Therefore, we have: 
	\begin{eqnarray} \label{GDOP_1}
	GDOP = \sqrt{\frac{\sigma^2_x+\sigma^2_y+\sigma^2_z}{\sigma^2_r}}.
	\end{eqnarray}
	
	Let $(x, y, z)$ denote the drone's position and $(x_i, y_i, z_i)$ denote the position for each of the ultrasound transmitter beacons in the room. Then, the drone range to each beacon is calculated from the following:
	\begin{eqnarray} \label{r}
	r_i = \sqrt{(x-x_i)^2 + (y-y_i)^2 + (z-z_i)^2}.
	\end{eqnarray}
	Because of the ranging measurement error, the exact $r_i$ is not known and that causes errors in the solution of Eq.~\ref{r} for
	$(x, y, z)$. To find a relationship between the solution errors and the ranging errors between the drone and each of the ultrasound transmitter beacons in the room, similar to \cite{xDOP_Formulas}, we take the differential of Eq.~\ref{r} and ignore terms beyond first order:
	\begin{eqnarray}
	\Delta r_i = \frac{\Delta x(x-x_i) + \Delta y(y-y_i) + \Delta z(z-z_i)}{\sqrt{(x-x_i)^2 + (y-y_i)^2 + (z-z_i)^2}} \nonumber \\
	= \Delta x \cos \alpha_i + \Delta y \cos \beta_i + \Delta z \cos \gamma_i \nonumber
	\end{eqnarray} 
	where $[\cos \alpha_i \ \cos \beta_i \ \cos \gamma_i]^T$ is the unit vector pointing from the drone to the $i$-th beacon.
	
	Let $\mathbf{\Delta X} = [\Delta x \ \Delta y \ \Delta z]^T$ be the position error vector and $\mathbf{\Delta R} = [\Delta r_1 \cdots \Delta r_n]^T$ be the target range error vector. Then we can define matrix $\textbf{U}$ as:
	\begin{eqnarray}
	\textbf{U} & = & \begin{bmatrix}
	u^1_1 & u^1_2 & u^1_3 \\
	\vdots     & \vdots     & \vdots\\
	u^n_1 & u^n_2 & u^n_3 \\
	\end{bmatrix}\nonumber
	\end{eqnarray}
	
	where $[u^i_1 \ u^i_2 \ u^i_3] = [\cos \alpha_i \cos \beta_i \cos \gamma_i]$. Now we can write $\mathbf{\Delta R} = \textbf{U} \mathbf{\Delta X}$ and then we have $\mathbf{\Delta X} = (\textbf{U}^T\textbf{U})^{-1} \textbf{U}^T \mathbf{\Delta R}$. We know that:
	\begin{eqnarray} \label{GDOP_2}
	\textbf{Cov}(\mathbf{\Delta X}) = \textbf{E}(\mathbf{\Delta X}\mathbf{\Delta X}^T) =   
	\begin{bmatrix}
	\sigma^2_x & \sigma_{xy} & \sigma_{xz} \\
	\sigma_{yx} & \sigma^2_y & \sigma_{yz}  \\
	\sigma_{zx} & \sigma_{zy} & \sigma^2_z  \\
	\end{bmatrix} 
	\end{eqnarray}
	If we assume that Var($r_i$) = $\sigma^2_r$ and that the errors $\Delta r_i$ are uncorrelated, then:
	\begin{eqnarray} \label{GDOP_3}
	\textbf{E}(\mathbf{\Delta X}\mathbf{\Delta X}^T) = \textbf{E}(((\textbf{U}^T\textbf{U})^{-1} \textbf{U}^T \mathbf{\Delta R})((\textbf{U}^T\textbf{U})^{-1} \textbf{U}^T \mathbf{\Delta R})^T) \nonumber \\
	= (\textbf{U}^T\textbf{U})^{-1} \textbf{U}^T \textbf{E}(\mathbf{\Delta R}\mathbf{\Delta R}^T) ((\textbf{U}^T\textbf{U})^{-1} \textbf{U}^T)^T \nonumber \\
	= (\textbf{U}^T\textbf{U})^{-1} \textbf{U}^T \textbf{U} (\textbf{U}\textbf{U}^T)^{-1} \sigma^2_r = (\textbf{U}^T\textbf{U})^{-1} \sigma^2_r \nonumber
	\end{eqnarray}
	Eq.~\ref{GDOP_1}, Eq.~\ref{GDOP_2} and the above result show that the diagonal elements of the $(\textbf{U}^T\textbf{U})^{-1}$ can be used to calculate the GDOP. GDOP consists of Vertical Dilution of Precision (VDOP) and Horizontal Dilution of Precision (HDOP), i.e., $GDOP = \sqrt{{HDOP}^2+{VDOP}^2}$ where HDOP represents the effect of the relative geometry between transmitters and the receiver on the $X-Y$ plane estimation accuracy and VDOP, on the other hand, shows the impact of geometry on the $Z$-axis estimation. This explains why we saw different accuracy for the $Z$-axis estimation and $X-Y$ plane estimation in our preliminary tests. Table \ref{table:GDOP} shows the evaluation of the GDOP values.
	
	\begin{table}
		\caption{Evaluation of GDOP Values}
		\label{table:GDOP}
		\begin{center}
			\begin{tabular}{ |c | c | }
				\hline
				\textbf{GDOP Values} & \textbf{Evaluation of the geometry of the beacons} \\
				\hline
				$<1$ & Measurements error or redundancy \\
				\hline
				$1$ & Ideal \\
				\hline
				$1-2$ & Very Good  \\
				\hline
				$2-5$ & Good \\
				\hline
				$5-10$ & Medium \\
				\hline
				$10-20$ & Sufficient  \\
				\hline
				$>20$ & Bad \\
				\hline
			\end{tabular}
		\end{center}
	\end{table}

	\subsection{Optimized Beacon Placement}
	Here, we propose an optimization algorithm to minimize the $Z$-axis estimation error induced by the relative geometry between transmitter beacons and the receiver, while keeping the horizontal error related to geometry in an acceptable range.
	
	Even though the optimal beacon placement for localization of a single static target in two-dimension scenarios is well-understood, the optimal placement for a mobile target in a three-dimensional space is still an open problem \cite{CMU_paper}. Finding an optimal beacon placement configuration for indoor localization to minimize the localization error due to the relative geometry between the transmitter beacons and the target receiver at any given position is a well-established NP-Hard problem \cite{Efficient_Beacon_Placement, CMU_paper, Novel_Beacon_Placement, BLE_Localization_Precision_Limits}. 
	
	Most of the earlier localization techniques have attempted to localize the unknown object just in two-dimension scenarios, which does not consider the real-world geometrical arrangement between the target and beacons in three dimensions. In the following, we will describe our systematic approach to find the optimized beacon placement in the room to improve the $Z$-axis estimation accuracy and mitigate the overall estimation error by developing a greedy algorithm.
	
	\subsubsection{Problem Formulation}
	Find the optimal placement for a set of four ultrasound transmitter beacons with the goal of minimizing the $VDOP_{avg}$ and keeping the $HDOP_{avg}$ below a required threshold, where $VDOP_{avg}$ and $HDOP_{avg}$ are the average of calculated VDOP and HDOP for a set of four beacons on all the given positions in the drone domain. Due to the constant mobility of the drone, it is not sufficient to compute the VDOP and the HDOP just for one position. Therefore, we considered all the possible locations the drone may pass by during its flight (i.e., all the points in the drone domain) and computed the average of the VDOP and the HDOP over all those possible locations and used them in our calculations. Moreover, if we simply constructed the optimization framework to minimize the average GDOP, then there would be no guarantee that it would improve the $Z$-axis estimation accuracy; hence we used the average of the VDOP and the HDOP values. The optimization problem can be formulated as follows.
	\begin{eqnarray}
	\min& \sum_{Drone Domain} VDOP \nonumber \\ 
	s.t.& \ HDOP_{avg} < h \nonumber
	\end{eqnarray}
	
	The minimization of the average VDOP is because the goal here is to find the optimized beacon placement to improve the drone's height estimation accuracy. The consideration for keeping the average HDOP below a threshold is to have a reasonable overall $3$D localization error due to the relative geometry, i.e., this constraint ensures that improving the $Z$-axis estimation is not with the cost of sacrificing the $X-Y$ plane estimation accuracy.
	
	The acquired inputs are as follows. The drone domain, set $D$, is a subspace of the room where the drone is allowed to fly. Optimization calculations are based on the average of $VDOP$ and $HDOP$ over all the points in this domain. The beacon domain, set $B$, is acceptable locations for beacons in the room. The entire ceiling and top half of all walls are acceptable candidates for the beacon locations. $HDOP_{avg}$ tolerance $(h)$ is a constraint which dictates $HDOP_{avg}$ be smaller than $h$ and $VDOP_{avg}$ tolerance $(v)$ is a constraint which dictates $VDOP_{avg}$ be smaller than $v$.
	
	As we discussed earlier, if each measurement has the same uncertainty with zero mean and unit variance and they are uncorrelated from each other, then the aforementioned HDOP and VDOP in the above steps can be derived from the diagonal elements of the matrix $\textbf{Q}$ as follows:
	\begin{eqnarray}
	\textbf{Q} = (\textbf{U}^T\textbf{U})^{-1} = 
	\begin{bmatrix}
	\sigma^2_x & \sigma_{xy} & \sigma_{xz} \\
	\sigma_{yx} & \sigma^2_y & \sigma_{yz} \\
	\sigma_{zx} & \sigma_{zy} & \sigma^2_z  \\
	\end{bmatrix}, \nonumber 
	\end{eqnarray}
	where $VDOP = \sqrt{\sigma^2_z}$, $HDOP = \sqrt{\sigma^2_x+\sigma^2_y}$, and
	\begin{eqnarray}
	\textbf{U} =   
	\begin{bmatrix}
	\frac{x_{1}-x}{r_1} & \frac{y_{1}-y}{r_1} & \frac{z_{1}-z}{r_1} \\
	\frac{x_{2}-x}{r_2} & \frac{y_{2}-y}{r_2} & \frac{z_{2}-z}{r_2} \\
	\frac{x_{3}-x}{r_3} & \frac{y_{3}-y}{r_3} & \frac{z_{3}-z}{r_3} \\
	\frac{x_{4}-x}{r_4} & \frac{y_{4}-y}{r_4} & \frac{z_{4}-z}{r_4} \\
	\end{bmatrix}, \nonumber 
	\end{eqnarray}
	where $(x,y,z)$ is the drone's position, $(x_i,y_i,z_i)$ is the location coordinate of the $i$-th ultrasound transmitter beacon, and $r_i$ represents the distance between the drone and the $i$-th ultrasound transmitter beacon.
	
	\subsubsection{Algorithm Design}
	In order to find a solution with manageable computational time and effort, we develop a greedy algorithm that is based on the class of Evolutionary Algorithms (EAs) to find the beacon placement.
	
	\begin{algorithm}\label{Alg}
		\caption{Beacon Placement Evolutionary Algorithm}
		\begin{algorithmic}[1]
			\newcommand\algorithmicinput{\textbf{Input:}}
			\newcommand\INPUT{\item[\algorithmicinput]}
			\INPUT Drone domain (D), Beacon domain (B), $HDOP_{avg}$ tolerance $(h)$, $VDOP_{avg}$ tolerance $(v)$
			\newcommand\algorithmicoutput{\textbf{Output:}}
			\newcommand\OUTPUT{\item[\algorithmicoutput]}
			\OUTPUT Desirable placement for a set of four beacons
			\WHILE{$VDOP_{avg} > v \ \& \ HDOP_{avg} > h$}
			\STATE Generate a set of P random individuals, where each individual is a set of four beacons 
			\FOR {$i = 1$ to $i = number \ of \ iteration$}
			\STATE Check the fitness of all available individuals;
			\STATE Kill the worst ones to keep having P individuals;
			\STATE Select the individuals with better fitness as Parents;
			\STATE Crossover each two adjacent parents and make a new offspring;
			\ENDFOR
			\ENDWHILE
		\end{algorithmic}
	\end{algorithm}
	
	In the algorithm, first, an initial set of $P = 50$ randomly-generated individuals is created. Each individual here is a set of four transmitter beacons selected randomly from the beacon domain. To avoid being trapped in a local minimal, we distributed the initial individuals in different groups: all the beacons on the ceiling, all of them on the walls, or some on the ceiling and some on the walls. We generated some random individuals for each of these groups. That is the reason we chose $50$ randomly-generated individuals, because if it was less than $50$, it was not able to include all the different groups, and if it was more than $50$, then it made the solving time unnecessarily longer. 
	
	Then, individuals are sorted according to the fitness (cost) function and those with better fitness are chosen for reproduction. The fitness function is the average of VDOP over the entire drone domain that is achieved using that specific arrangement of four beacons. Then, the algorithm picks the first $40$ individuals in the line as a parents group that will be used to reproduce new individuals. Every two adjacent of them make a new set of four beacons using a crossover technique; therefore, it would be $20$ off-springs total. Next, the algorithm checks every new-generation individual according to the fitness function. Out of $70$ total populations, including both the parents and off-springs, the last $20$ in the line will be eliminated, so the population remains $50$. 
	
	For generating a new off-spring, each set of parents includes $8$ beacons total ($4$ beacons per parent). The crossover technique switches some of the coordinate parameters of the first four beacons with the ones from the second set of four beacons.
	
	After the initialization, the described procedure would repeat for $100$ iterations. We checked the algorithm for different iteration numbers and saw that larger ones (e.g., $1000$) just make the process slower without bringing any significant improvement on the final result. Moreover, iterations less than $100$ still did not provide a minimal answer, so that was why we picked $100$ as the number of iterations. After that, the first individual in the line according to the fitness function is selected. If it has an average VDOP and HDOP over the entire drone domain less than $v$ and $h$ respectively, then that individual represents the final answer which is the beacon placement configuration in the room. Otherwise, the algorithm starts over again, from generating the $50$ first new random individuals and repeating the procedure. The algorithm terminates whenever the final result satisfies the constraints.
	
	\subsection{Additional Sensor for Height Estimation}
\begin{figure}
		\centering
		\begin{subfigure}{.5\textwidth}
			\centering
			\includegraphics[width=\textwidth]{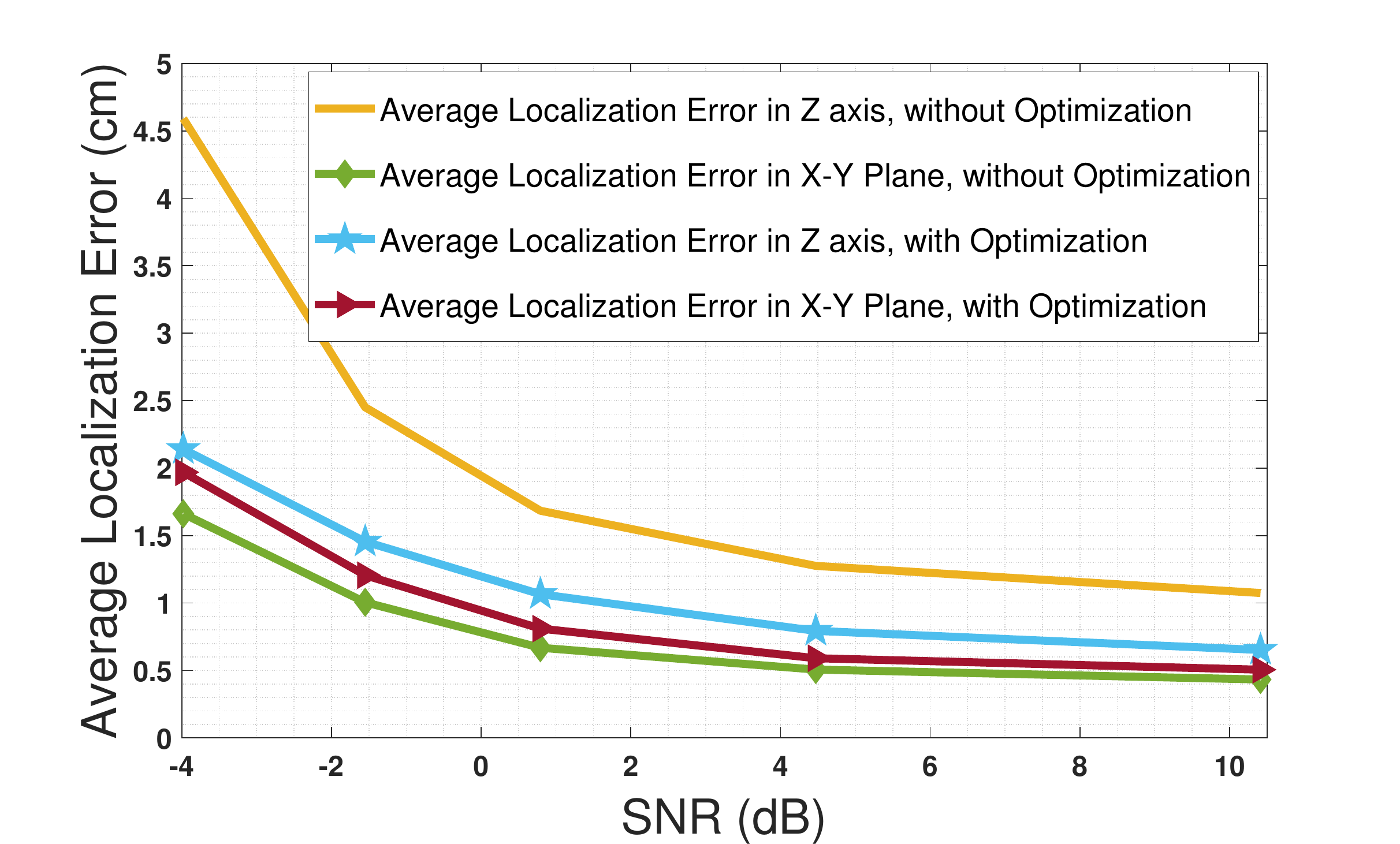}
			\caption{}
			\label{SNR_comp}
		\end{subfigure}
		\begin{subfigure}{.5\textwidth}
			\centering
			\includegraphics[width=\textwidth]{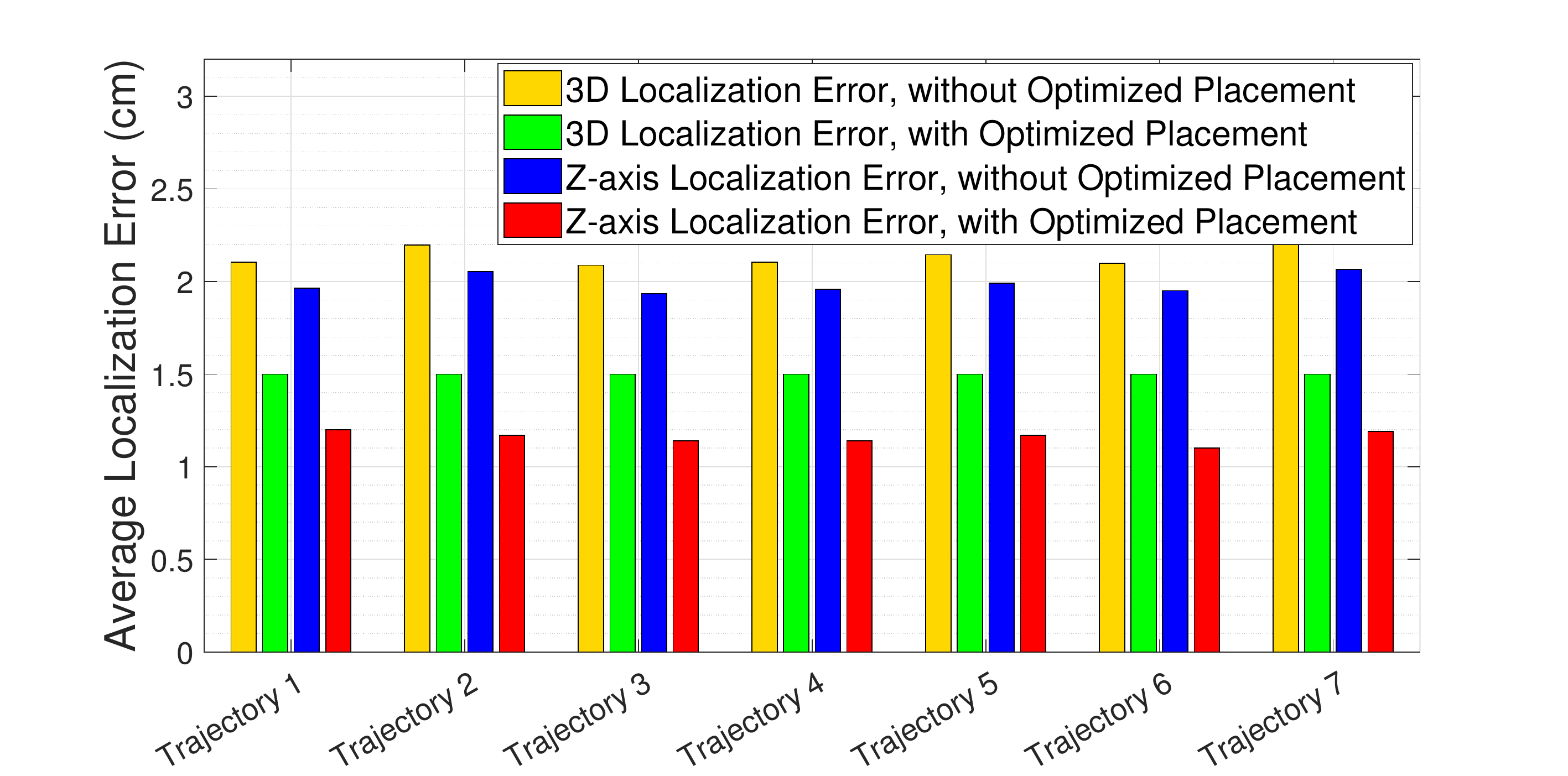}
			\caption{}
			\label{Trajectories}
		\end{subfigure}
		\caption{(a) Average Localization error vs. SNR (dB). (b) Z-axis and overall three-dimensional localization error for seven random trajectories before and after applying the beacon placement optimization algorithm.}
		\label{}
	\end{figure}

	We have improved the accuracy of the $Z$-axis estimation by optimizing the beacon placement. However, as is seen in Fig.~\ref{SNR_comp}, still the $Z$-axis estimation accuracy is slightly worse than the location estimation accuracy of the $X-Y$ plane. To further improve the height estimation, we use a separate ultrasonic transceiver mounted on-board the drone to continuously estimate the height. Then, using a filter, we incorporate this measurement with the $Z$-axis estimation that has been already available from the first step of iDROP. This significantly improves the $Z$-axis estimation accuracy.
	
	We mount the ultrasonic transceiver facing upward on-board the drone and find the distance between the drone and the ceiling by calculating the time of flight of the ultrasonic signal transmitted from the sensor on-board the drone, after it is reflected from the ceiling. Then, simply by subtracting this result from the room height, we find the drone's height at each moment. The channel between the drone and the ceiling is usually more reliable than the one between the drone and the floor because usually there are no objects between the drone and the ceiling that induce errors. The height estimation using this extra ultrasonic transceiver is then: $\ h_{drone} = \ H \ - \ d$ and $d = \ c_{sound}\cdot t/2$;
	where $d$ is the distance between the drone and the ceiling, $t$ is the total time that takes the signal to travel from ultrasonic transceiver on-board the drone and hit the ceiling, reflecting, and is received in the ultrasonic transceiver on-board the drone, $H$ is the room height, and $h_{drone}$ is the estimation for drone's height. The final revised height estimation is then: $z_{revised} = w_1\cdot z + w_2\cdot h_{drone}$, where $w_1$ and $w_2$ are the weights assigned for the $Z$-axis estimation from the first stage of iDROP and this stage, and $z$ is the estimated height from the first stage of the iDROP.
	
	\section{Experimental Results and Evaluations} \label{Experimental Tests and Results}
	\subsection{Experimental Setup}
\begin{figure}
		\centering
		\includegraphics[height=1.4in,width=2.65in,trim={3cm 21cm 19cm 1cm},clip]{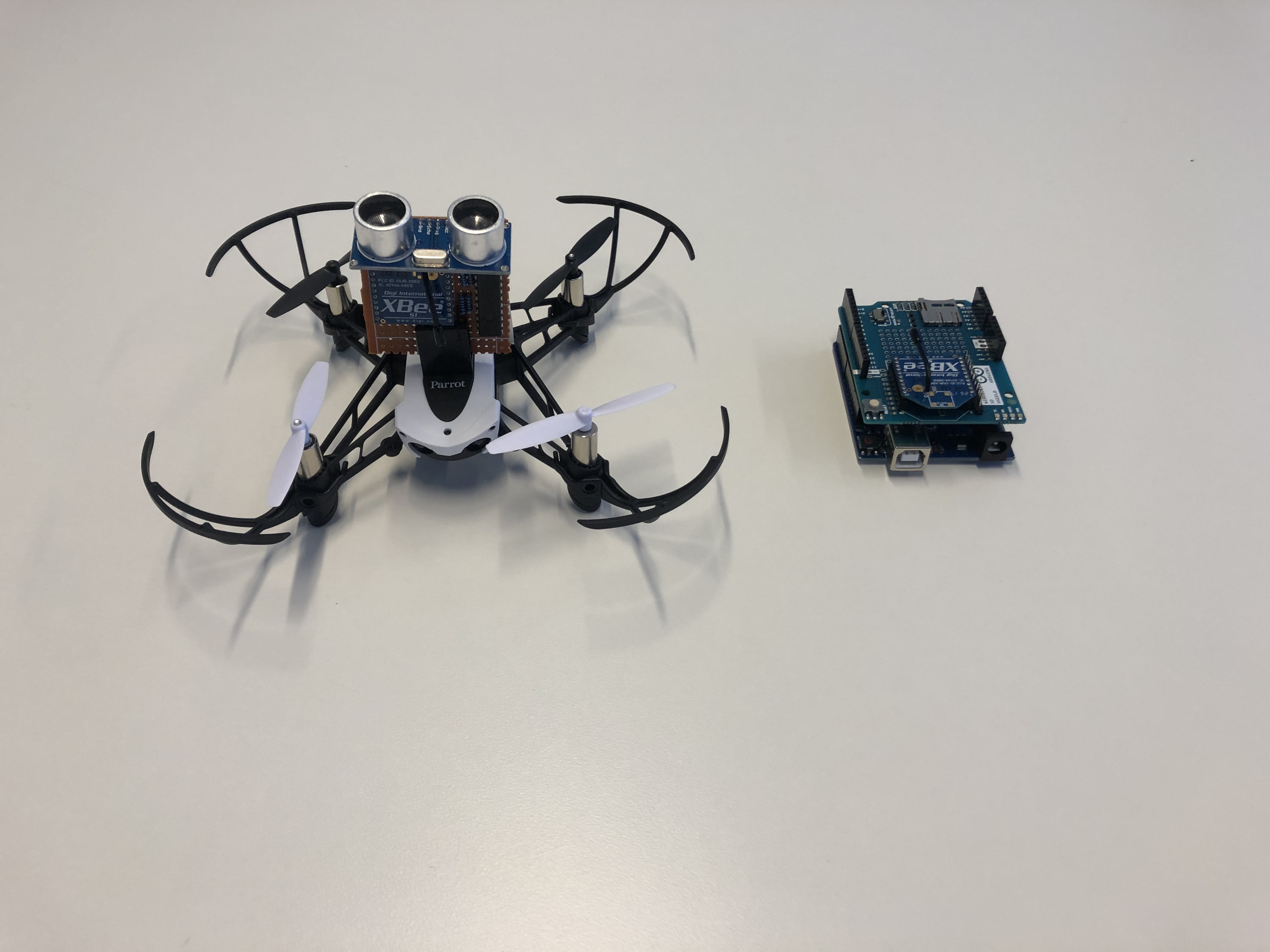} 
		\caption{Parrot Mambo Drone equipped with the ultrasound transceiver system at the left and receiver side on the right.}
		\label{testbed}
	\end{figure}

	To assess the performance of iDROP, we conducted some experimental tests coupled with MATLAB simulations. As is seen in Fig.~\ref{testbed}, the experimental test setup consists of two stations: first, the drone and the system on-board it, and the second one is the ground control station which helps to input the transmitted data into the MATLAB program running on a Dell XPS $15$ laptop. The drone that we used for the experiment is a Parrot Mambo Drone. It is a cheap, off-the-shelf, and ultralight drone suitable for indoor experiments. Also, it has the capability of carrying some light loads. The designed system mounted on-board the drone consists of an Arduino Uno micro-controller connected to an $HC$-$SR04$ sensor for ultrasonic distance measurement purposes and a XBee S1 module for wireless communication with the ground controller. In the ground control unit, another Arduino Uno micro-controller connected to a XBee S1 receives the data and further transfers it into the MATLAB program running on the laptop. All the experiments were conducted in a hallway inside a building with dimensions $5$~m$~\times$~$5$~m~$\times$~$4$~m. This hallway contains several objects and obstacles, including tables, chairs, glass windows, etc.
	
	\begin{table*} [t]
		\caption{Comparison of iDROP with Comparable Schemes.}
		\label{table:Comparison}
		\begin{center}
			\begin{tabu} to \textwidth {XXXXXXX}
				\toprule
				\textbf{System} & \centering \textbf{Measurement} & \centering \textbf{Multi-path Solution} & \centering \textbf{$3$D Localization} & \centering \textbf{Claimed Accuracy} & \centering \textbf{Optimized Beacon Placement} & \centering \textbf{Improved $Z$-axis Estimation}\\
				\hline
				\cite{Spread_Spectrum_and_MEMS} $(2014)$ & \centering Ultrasound, TOA & \centering None & \centering No & \centering $2$~cm for $2$D & \centering No & \centering No \\
				\hline
				\cite{MobiSys_Follow_Me_Drone} $(2017)$ & \centering Acoustic, TOA & \centering FMCW & \centering No & \centering $2.6$~cm for $1$D & \centering No & \centering No \\
				\hline
				\cite{Ultrasonic_Quadrotor_2019} $(2019)$ & \centering Ultrasound, TOA  & \centering None & \centering Yes & \centering $5.2$~cm for $3$D & \centering No & \centering Yes  \\
				\hline
				\cite{ROLATIN} $(2020)$ & \centering Ultrasound, TOA & \centering FHSS & \centering Yes & \centering $1.4$~cm for $3$D & \centering No & \centering No \\
				\hline
				\cite{RAIL} $(2021)$ & \centering Ultrasound, TOA & \centering FH-CDMA & \centering Yes & \centering $1.5$~cm for $3$D & \centering No & \centering No \\
				\hline
				\textbf{iDROP} & \centering \textbf{Ultrasound, TOA} & \centering \textbf{FH-CDMA} & \centering \textbf{Yes} & \centering \textbf{$1.2$~cm for $3$D} & \centering \textbf{Yes} & \centering \textbf{Yes} \\
				\bottomrule
			\end{tabu}
		\end{center}
		
	\end{table*}
	
	\subsection{Final Results}
	
	In Fig.~\ref{SNR_comp}, a comparison of the localization error for both the $Z$-axis and the $X-Y$ plane, in different SNRs, before and after using the proposed EA optimization framework is seen. The new $(x,y,z)$ coordination of the ultrasound transmitters in the room that is obtained from the EA optimization framework is $(4.5,0,2.5)$, $(5,4,3.5)$, $(1,5,2)$, and $(1.5,2,4)$ where all the numbers are in meter. As is seen in this figure, using the optimized placement for transmitters improves the $Z$-axis localization accuracy significantly. Moreover, it does not drastically degrade the $X-Y$ plane localization accuracy. The reason for developing the EA optimization framework is to find the optimal placement of transmitters in the room to mitigate the localization error due to the relative geometry between the transmitters and the target drone. More specifically, it is designed in a way to mitigate the $Z$-axis localization error without significantly degrading the $X-Y$ plane localization accuracy. This plot shows that the proposed EA optimization framework performs as is expected.
	
	In Fig.~\ref{Trajectories}, the $Z$-axis localization error for seven different random drone trajectories in cases where the hybrid FH-CDMA technique is used for localization in comparison with the cases where leverage the hybrid FH-CDMA technique for localization in combination with using the new optimal coordination for transmitters, is seen. Moreover, this figure shows a similar comparison for the overall three-dimensional localization error. As depicted in this figure, for all of these random trajectories for the drone in the room, both the $Z$-axis and the overall localization error improve when the optimized beacon placement is used.
	
	\begin{figure}
		\centering
		\begin{subfigure}{.5\textwidth}
			\centering
			\includegraphics[width=\textwidth]{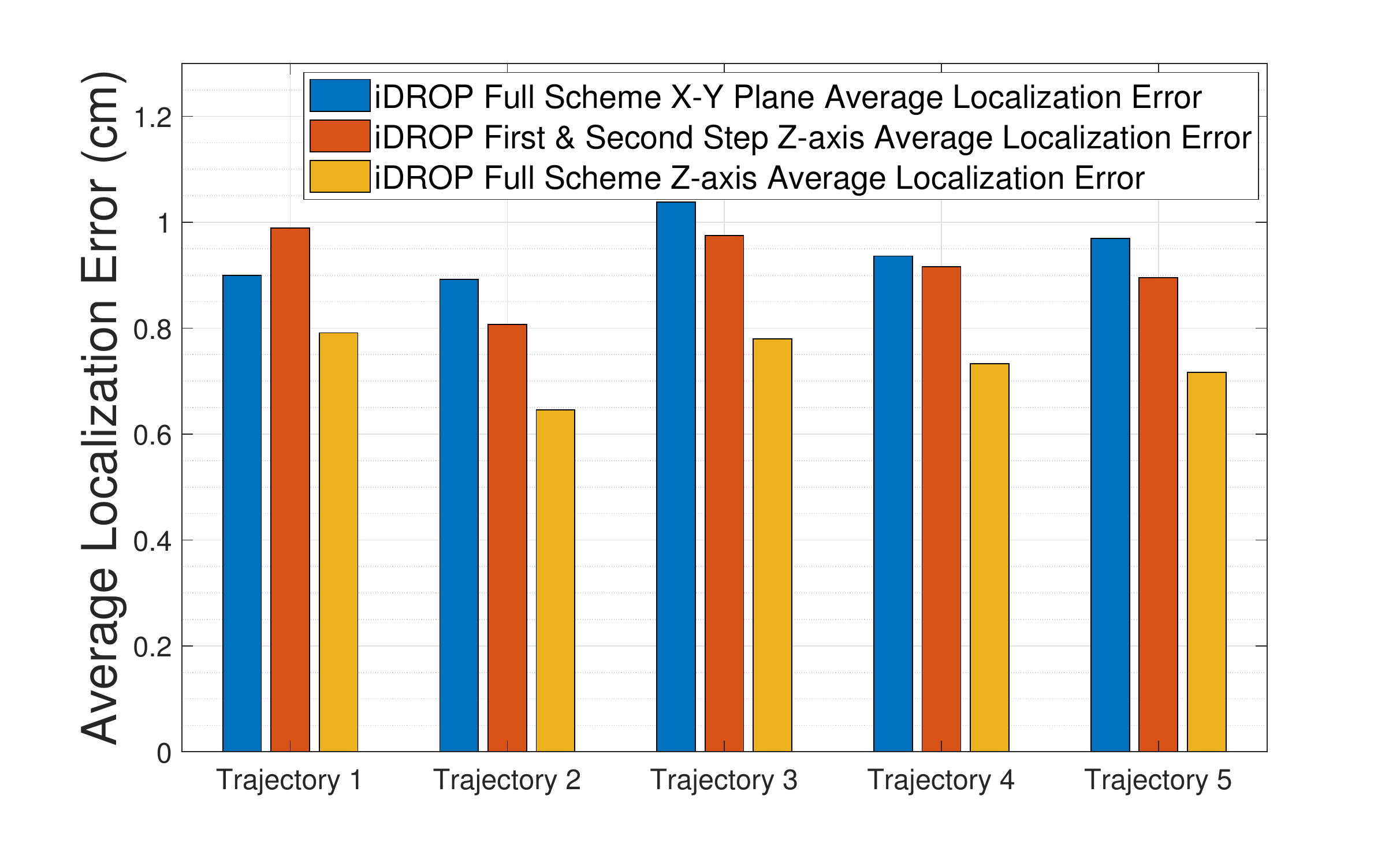}
			\caption{}
			\label{Experiment_XYvsZ}
		\end{subfigure}
		\begin{subfigure}{.5\textwidth}
			\centering
			\includegraphics[width=\textwidth]{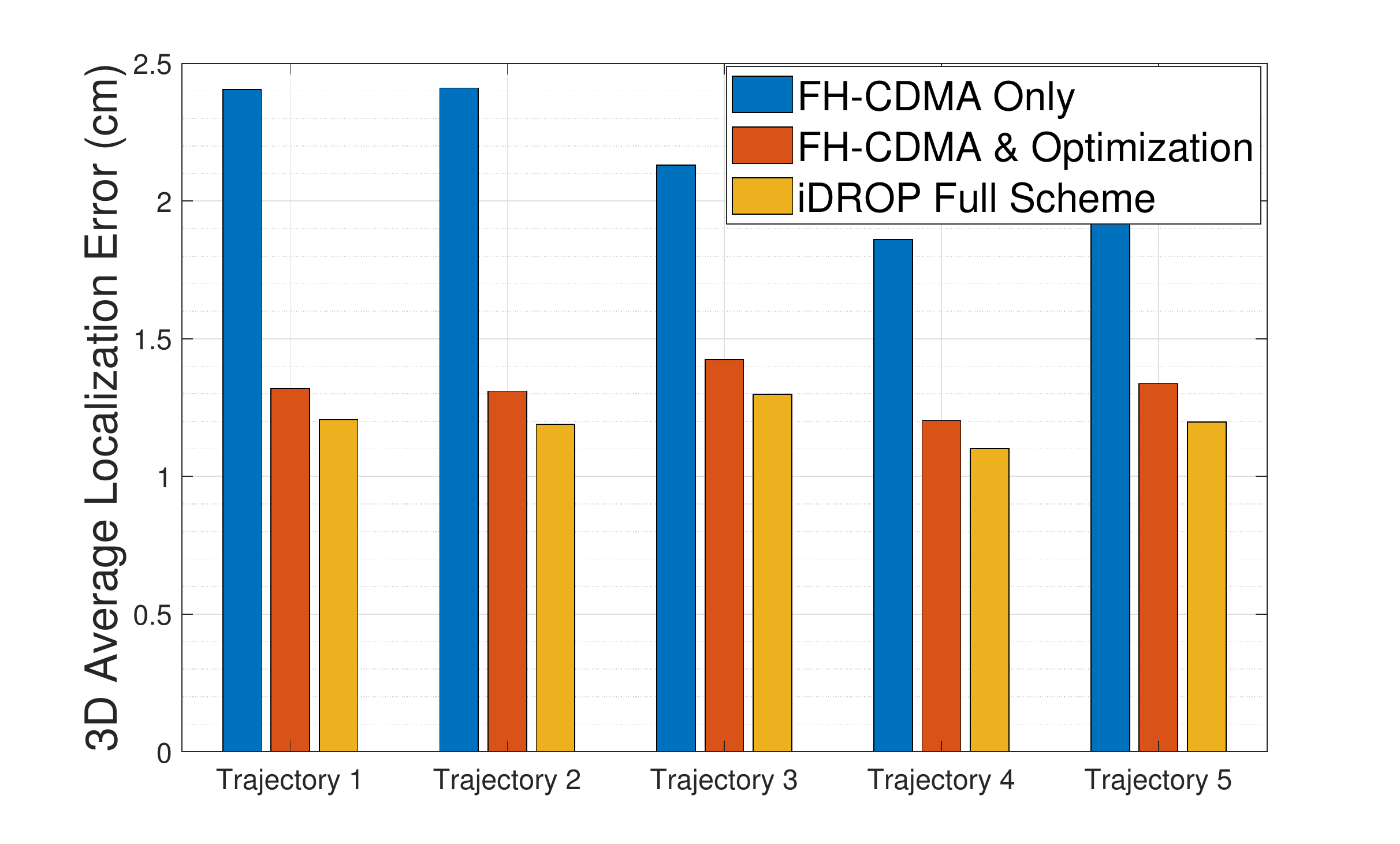}
			\caption{}
			\label{Experiment_Overall}
		\end{subfigure}
		\caption{(a) Evaluating the performance of iDROP: Comparison between the $X-Y$ plane average estimation error and the $Z$-axis. (b) Evaluating the performance of iDROP: overall three-dimensional localization accuracy.}
		\label{}
	\end{figure}
	
	In Fig.~\ref{Experiment_XYvsZ}, the localization error of the $Z$-axis with the $X-Y$ plane is compared. This figure justifies the necessity of having the auxiliary sensor for height estimation. As is seen in this figure, even though the optimized placement for transmitters improved the $Z$-axis localization accuracy significantly, the $Z$-axis error may not be as low as the $X-Y$ plane localization error. This figure shows how the last step of iDROP further improves the $Z$-axis estimation by constantly transferring the measured data from the ultrasound sensor on-board the drone ($HC$-$SR04$) to the receiver module connected to the Dell XPS $15$ laptop. Therefore, iDROP successfully improves the $Z$-axis estimation without sacrificing the $X-Y$ Plane localization accuracy.
	
	In Fig.~\ref{Experiment_Overall}, the performance of iDROP with that of the benchmark scheme (which relies only on FH-CDMA distance estimation to localize a target drone) in terms of the overall three-dimensional localization error is compared. The average value of three-dimensional localization error for iDROP is $1.2$~cm. As is seen in the figure, the benchmark scheme’s localization error is more than twice that of iDROP. This is because the benchmark scheme merely focuses on mitigating ranging-based error by deploying the FH-CDMA communication scheme for localization. Other drone localization schemes proposed in the literature do the same and try to improve the localization accuracy by proposing their technique to mitigate the ranging-based error. However, iDROP proposes a scheme that deals with both ranging-based error and geometry-related error and in this way, it further improves the accuracy.
	
	iDROP achieves significant improvement in comparison with other drone localization schemes in the literature \cite{ROLATIN, Robust_Broadband, Ultrasonic_Quadrotor_2019, Spread_Spectrum_and_MEMS, Spread_Spectrum_Ultrasound_and_Time-of-Flight_Cameras}. For instance, in \cite{Ultrasonic_Quadrotor_2019}, O'Keefe et al. proposed a scheme that incurs an average error of $5.2$~cm for three-dimensional localization for drones. The scheme proposed by Segers et al. \cite{Spread_Spectrum_and_MEMS} incurs an error of $2$~cm or greater in terms of localization error just for the $X-Y$ plane (two-dimensional localization). Table~\ref{table:Comparison} compares iDROP with other comparable schemes introduced in the literature.
	
	As is seen in Table~\ref{table:Comparison}, iDROP proposes a novel solution for each of the existing challenges, and that is why it has a better overall performance in comparison with the works in the literature. For instance, in \cite{MobiSys_Follow_Me_Drone}, Mao et al. proposed an FMCW technique to overcome the multi-path fading effect; however, their scheme was designed just for tracking a drone on a line and they did not consider three-dimensional localization and its challenges. In \cite{ROLATIN}, Famili et al. proposed a multi-path robust scheme for three-dimensional localization of drones; however, their scheme had significantly low accuracy for $Z$-axis estimation. Moreover, their claimed accuracy is merely based on MATLAB simulations and they did not provide any real-life experiments with an actual drone to assess their proposed scheme. In \cite{Ultrasonic_Quadrotor_2019}, O'Keefe failed to propose an optimal solution for beacons because of their choice of system design and lack of signal separation capability. They also did not consider a multi-path-robust communication technique. Therefore, their proposed scheme had high localization errors. In another work \cite{RAIL}, Famili et al. proposed a multi-path robust system for drones' three-dimensional localization in indoor environments. Even though their proposed scheme fixed the signal separation challenge and eliminated the unnecessary communication link between the drone and the transmitter beacons in the room, they failed to explain the reason behind having a bad $Z$-axis estimation and their system lacked the optimal beacon placement analysis. iDROP has robustness against multi-path fading and noise and provides signal separation capability. Moreover, it proposes an optimized placement for beacons in the room and improves the $Z$-axis estimation accuracy. Overall, iDROP provides a highly accurate $3$D localization in real-time scenarios for drones.
	
	\section{Conclusions} \label{Conclusion}
	In this paper, we presented a multi-path-robust localization system with optimized beacon placement that can be used for autonomous navigation of drones in indoor environments. First, the drone can locate itself in three-dimensional space using the TOA of the received FH-CDMA ultrasound waveform. Then, to improve the $Z$-axis estimation accuracy, an EA optimization framework finds an optimized location for the transmitter beacons in the room. Moreover, an additional ultrasound transceiver provides a separate measurement of the drone's height for improving the $Z$-axis estimation accuracy. Finally, we evaluate the performance of our proposed system by conducting experiments coupled with simulations.
	
	\bibliographystyle{ACM-Reference-Format}
	\bibliography{ref_iDROP_Revision}
		
\end{document}